\documentclass[3p,times]{article}

\usepackage{ulem}
 \usepackage{cite}

\usepackage{amssymb}
\usepackage{xspace}
\usepackage[colorinlistoftodos]{todonotes}
\usepackage{multicol}
\usepackage{graphicx}
\usepackage{lipsum}

\usepackage{lineno}

\usepackage[figuresright]{rotating}

\newcommand{\Nevt}{\mbox{$N_{\mathrm{evt}}$}\xspace}

\newcommand{\sqn}{\mbox{$\sqrt{s_{\mathrm{NN}}}$}\xspace}

\newcommand{\energy}{\mbox{\sqn=200\,GeV}\xspace}

\newcommand{\AuAu}{\mbox{Au+Au}\xspace}

\newcommand{\Npart}{\mbox{$N_{\mathrm{part}}$}\xspace}

\newcommand{\nf}{\mbox{$N_{\mathrm{f.n.}}$}\xspace}
\newcommand{\Npt}{\mbox{$N_{\mathrm{part}}^{\mathrm{true}}$}\xspace}
\newcommand{\Npr}{\mbox{$N_{\mathrm{part}}^{\mathrm{rec}}$}\xspace}

\newcommand{\avgNpart}{\mbox{$\langle\Npart\rangle$}\xspace}

\newcommand{\af}{\mbox{$A_{f}$}\xspace}
\newcommand{\zf}{\mbox{$Z_{f}$}\xspace}
\newcommand{\az}{\mbox{\af/\zf}\xspace}
\newcommand{\pz}{\mbox{$p_{z}$}\xspace}
\newcommand{\pf}{\mbox{$p^{F}$}\xspace}

\newcommand{\gD}{\mbox{DPMJet}\xspace}
\newcommand{\gQ}{\mbox{QGSM}\xspace}

\begin{document}





\begin{center}
{\large \bf A Centrality Detector Concept}
\vspace{5mm}


Sourav Tarafdar, Zvi Citron and Alexander Milov\\
\vspace{5mm}

{\it \footnotesize Department of Particle Physics and Astrophysics, Weizmann Institute of Science, \\
234 Herzl str., Rehovot 7610001, Israel}
\end{center}

\author{}


\begin{abstract}
The nucleus-nucleus impact parameter and collision geometry of a heavy ion collision are typically characterized by assigning a collision ``centrality". 
In all present heavy ion experiments centrality is measured indirectly, by detecting the number of particles or the energy of the particles produced in the interactions, typically at high rapidity. Centrality parameters are associated to the measured detector response using the Glauber model. This approach suffers from systematic uncertainties related to the assumptions about the particle production mechanism and limitations of the Glauber model. In the collider based experiments there is a unique possibility to measure centrality parameters by registering spectator fragments remaining from the collision. This approach does not require model assumptions and relies on the fact that spectators and participants are related via the total number of nucleons in the colliding species. This article describes the concept of a centrality detector for heavy ion experiment, which measures the total mass number of all fragments by measuring their deflection in the magnetic field of the collider elements. 
\end{abstract}






\section{Introduction}
\label{sec:intro}
The field of relativistic heavy ion (HI) collisions is a rapidly developing branch of modern nuclear physics whose goal is to study  the nature of the strong force. An extensive scientific program carried out by several experimental collaborations at the Super Proton Synchrotron (SPS) at CERN, the Relativistic Heavy Ion Collider (RHIC) at BNL, and recently at the Large Hadron Collider (LHC) at CERN has charted the creation of hot, dense, thermalized QCD medium.  The study of this medium reveals properties consistent with a Quark Gluon Plasma \cite{Shuryak200564,Jacak:2012dx}. An accurate and quantitative description of these properties is key to understanding the underlying physics of strong force interactions. 

The collisional geometry of HI interactions plays a very important role in defining the physics of the collision, and it is therefore crucial to characterize it with high precision. 
Ideally, the impact parameter ($b_{\mathrm{imp}}$) of the collision, the distance between the centers of colliding ions, would be used to define the collision centrality.  However $b_{\mathrm{imp}}$ can not be directly measured.   
The number of nucleons participating in the collision, (\Npart) i.e. the number of nucleons in both ions suffering at least one interaction with a nucleon of the counterpart ion, serves as a more experimentally accessible ordering parameter in defining centrality. \Npart is directly associated with the bulk particle production measured in HI collisions. In the ion fragmentation direction the number of charged particles and the energy they carry, is found to be proportional to \Npart~\cite{phobos}. The ``wounded nucleon model"~\cite{bialas} assumes the proportionality to be linear and accurately describes the experimental data at the SPS~\cite{Antinori:2004sz, PhysRevC.66.054902}. However, with increased energy, and considering mid-rapidity particle production the linearity is violated.  The ``number of participant quark" model appears to be a more complete description of the underlying processes \cite{PhysRevC.67.064905, EUJPhys2007, PhysRevC.71.024903, Quark_part_model}.  

Extracting \Npart from the response of the detector, typically located at forward rapidity on both sides of the interaction point, varies depending on the design of the experiment and the discretion of the collaboration. It is typically based on a Monte Carlo (MC) Glauber model ~\cite{phobos_mc} and involves simulating the particle production in the forward rapidity region and the detector response. Event centrality is defined by considering the distribution of the observed bulk particle production ($N_{\mathrm{ch}}$) in measured events. The $\mathrm{d}N/\mathrm{d}N_{\mathrm{ch}}$ distribution is divided into percentile classes, with the convention that the X\% of events with the largest $N_{\mathrm{ch}}$ are the most central events referred to as 0--X\% centrality. A similar classification is made in the simulation and thus an empirically defined class of events is related to \Npart. Typical systematic uncertainties on the determination of the \Npart vary from 1--2\% in the most central collisions to more than 10\% in more peripheral events.

A very interesting topic in HI physics is the study of asymmetric collisions systems such as the $p$+Pb at the LHC and $d$+Au at RHIC. Centrality determination in these systems is even more challenging than in symmetric system. Recent results from the ATLAS collaboration~\cite{multi} show that the approach based on the Glauber model which is used in the field for more than a decade may need improvement. Due to the very important role that centrality plays in the HI studies, improving the centrality determination should have large impact on the entire field of HI physics.

The main disadvantage of the presently used centrality determination approach is its use of model-based assumptions to relate the measured detector response to \Npart. 
Another disadvantage of the current method is its reliance on using the particles produced in the collision.  This often results in an intrinsic correlation between particles being measured as a function of centrality and the definition of centrality itself. 


In the collider based HI experiments there is a unique opportunity to measure \Npart by measuring ``spectator fragments", ions, protons and neutrons which continue to propagate in the same direction as the 
colliding ions 
before the interaction. Spectator fragments are formed by the nucleons which suffer no strong interaction with nucleons of the counterpart ion. The exact process of forming spectator fragments is not thoroughly studied, however the relation between \Npart and the the number of nucleons remaining in the fragments does not depend on this process 
\begin{equation}
\Npart = 2A - \sum_{i} A_{f}^{i}\hspace{2mm}.
\label{eq:npart}
\end{equation}
where $A$ is the mass number of a colliding ion and \af is the mass number of the spectator fragment. The sum is taken over all spectator fragments on both sides of the interaction point. Since nucleons forming fragments did not suffer strong interaction they retain full longitudinal momentum, \pz, and their momentum vector after interaction is approximately collinear with the vector of the initial ion. The trajectories of the particles in the collider depend on their mass-to-charge ($m/q$) ratio. Colliding ions with a particular $m/q$ stay on an equilibrium orbit, but fragments deviate from it, depending on their ratio $\az\propto m/q$. Since lighter nuclei have less neutrons compared to protons than heavier nuclei, lighter fragments formed after the collision typically have smaller mass-to-charge ratio. They are thus separated from the equilibrium beam by the magnetic structure of the collider according to their \az. This presents a unique opportunity for a collider-based HI experiment to build a centrality detector which measures \Npart by detecting spectator fragments and measuring their \af. Such approach is free of the main disadvantages present in the currently used centrality determination: it is not model dependent and it uses particles created by physics process which is decoupled from the particle production mechanism in the HI collision. This paper describes basic parameters of a centrality detector using the magnetic structure of RHIC. Three detector stations are considered on each side of the interaction point. The Zero Degree Calorimeters (ZDC)~\cite{zdc}, which are  existing integrated parts of operating RHIC experiments are used to detect free spectator neutrons. The main physics processes affecting detector performance are discussed based on the spectator fragmentation modelled using the \gD~\cite{DPMJET-III} and \gQ~\cite{QGSM_1,QGSM_2} event generators. 

Measuring the parameters of collisions by detecting the products remaining after the interaction was suggested in~\cite{Chwastowski:1995mi}. The NA49 experiment at the SPS measured the distribution of different fragments remaining after the interaction of a Pb ion with  a fixed Pb target~\cite{na49:spect}. At electron colliders, energy lost by the electron and positron was measured via their deflections in the magnetic structure of the collider rings~\cite{Aulchenko1996360}. This paper proposes an application of a similar approach to the HI collider experiments.

The paper is organised in the following sections:
Section~\ref{sec:tracing} calculates fragment trajectories in the magnetic structure of RHIC and explains the factors affecting the choice of detector station positions. Physics processes affecting distribution of the fragments on the surface of detector stations are discussed in section~\ref{sec:frag_gen}. Detector performance parameters, such as efficiency and centrality determination accuracy are presented in section~\ref{sec:res_main}.
   
\section{Modelling the collider structure}
\label{sec:tracing}
Spectator fragments with different \az are traced using the MAD-X (Methodical Accelerator Design) code~\cite{madx-web}.
In a formalism commonly used to design accelerators, the transport of particles from the interaction point (IP) to a given location $s$ along the ring can be described with a matrix:
\begin{eqnarray}
\left| \begin{array}{c}
x \\
x^{\prime} \\
y \\
y^{\prime} \\
z \\
\Delta p_{z}/p_{z} \\
\end{array} \right|
= \left( \begin{array}{cccccc}
a_{1,1} & a_{1,2} & & & a_{1,5} & a_{1,6} \\
a_{2,1} & a_{2,2} & & & & a_{2,6} \\
& & a_{3,3} & a_{3,4} & a_{3,5} & a_{3,6} \\
& & a_{4,3} & a_{4,4} & & a_{4,6} \\
& & & & 1 & a_{5,6} \\
& & & &   & 1 
\end{array} \right) \times
\left| \begin{array}{c}
x \\
x^{\prime} \\
y \\
y^{\prime} \\
z \\
\Delta p_{z}/p_{z}
\end{array} \right|_{IP}.
\label{eq:matrix}
\end{eqnarray}
 where $x,y$ are linear and $x^{\prime},y^{\prime}$ are angular transverse particle coordinates, $z$ is the longitudinal coordinate, and $\Delta p_{z}/p_{z}$ is the residual particle momentum.  All coordinates, including  $\Delta p_{z}/p_{z}$, are defined with respect to a particle in equilibrium orbit located at the center of the beam. Matrix elements which are considered equal to zero are not printed in the equation. 

The block-diagonal form of the matrix for indices $i,j\leq4$ corresponds to the case when the particle translations along $x$ and $y$ coordinates are decoupled from each other. Such approximation is sufficient for relatively short distances $s$ considered further. Particle coordinates in longitudinal directions affect translation in both $x$ and $y$ directions. At RHIC, the interaction region has a typical width of about 15\,cm around the nominal IP, and the event vertex position in $z$ can be measured with high precision for each collision. Measuring the vertex position eliminates the impact of $z$ coordinate in Eq.~\ref{eq:matrix} and therefore all interactions are modelled at the nominal IP.

Measuring spectators fragments relies on the fact that the particles with non-equilibrium longitudinal momentum $\Delta p_{z}/p_{z}\neq0$ have different trajectories in the collider. Coupling of the transverses coordinates to the longitudinal momentum is given by the last column of the matrix in eq.~\ref{eq:matrix}. For values of $\Delta p_{z}/p_{z}\ll1$ matrix elements $a_{i,j}$ can be considered as constant coefficients. However, for light fragments $\Delta p_{z}/{p_z}\approx 1-(\az)/(A_{\mathrm{Au}}/Z_{\mathrm{Au}})$ is significantly different from zero, and therefore a different matrix was calculated for each value of \az. 

\begin{figure}[!htb]
\begin{center}
\includegraphics[width=0.6\textwidth] {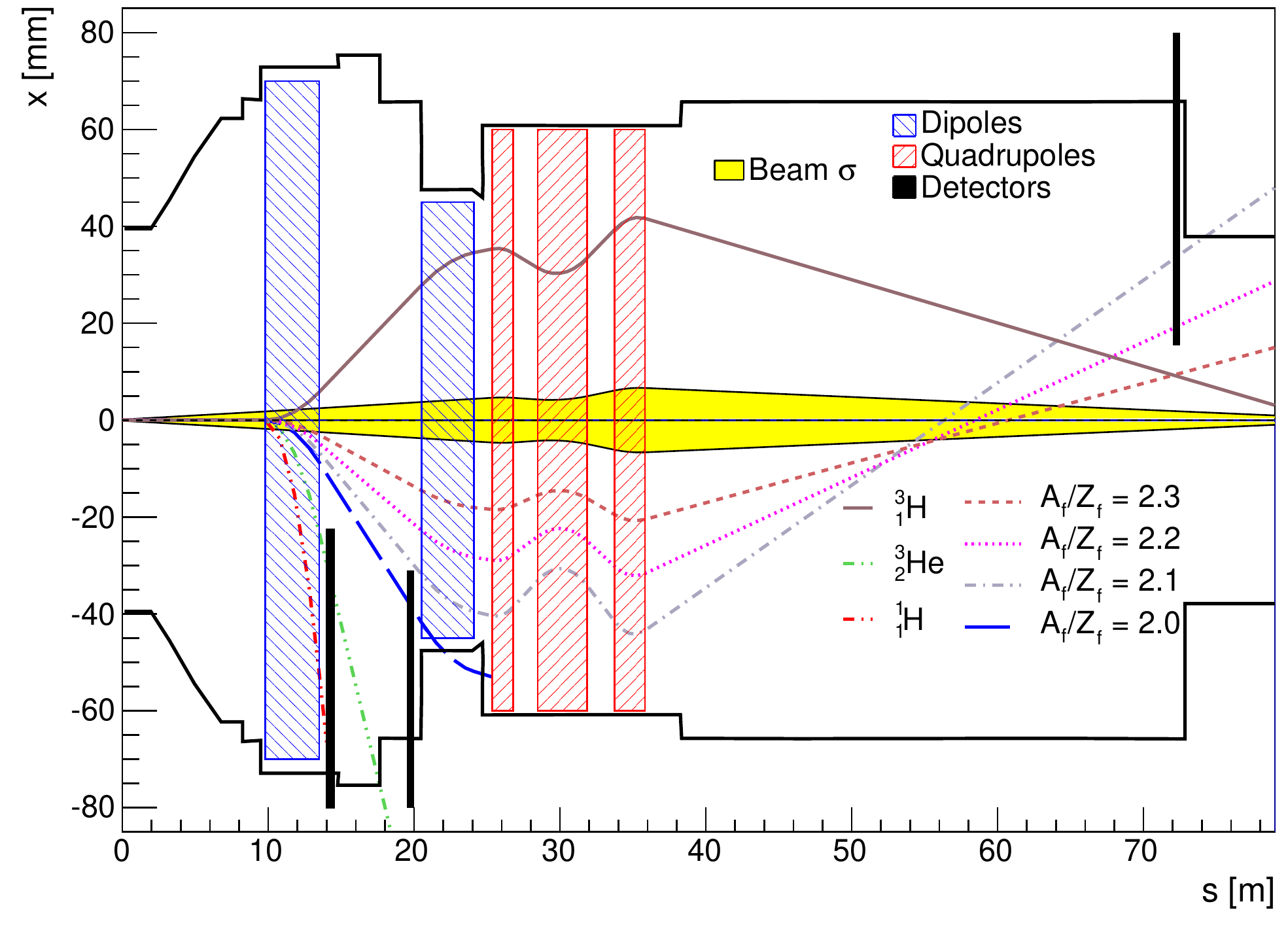}
\caption{Spectator fragment trajectories in the RHIC magnetic structure. Hashed boxes correspond to the locations of collider magnetic elements, the beam pipe is shown with the solid outer line, vertical lines indicate the locations of detector stations. The filled area is the size ($\sigma$) of the equilibrium beam. Lines are the trajectories of fragments with different \az.} 
\label{fig:tracing}
\end{center}
\end{figure}

Figure~\ref{fig:tracing} shows the spectator fragment trajectories calculated in $x$ coordinate.
The equilibrium Au beam with $A/Z=2.5$ is shown as the filled area. The size of the beam is $\sigma=\sqrt{\beta\varepsilon}$, where $\beta$ is the collider beta-function and $\varepsilon$ is the beam emittance.  At \energy  $\varepsilon=0.023$\,mrad$\times$mm and the value of the $\beta$ function at the IP is $\beta^{\star}=1.0$\,m.  Hashed areas correspond to the locations of collider magnetic elements, and the outer line shows the dimension of the vacuum beam pipe. The trajectories of fragments with $\az<2.5$ are shown in the overlaid lines. 
Neutron trajectories are not shown in the plot; they continue along straight lines in the laboratory coordinate system and terminate in the ZDC, located at a distance of 18\,m from the IP, between the first and second magnets. All particles are deflected to the same side of the Au beam, except tritium which has $\az=3$ and therefore appears on the other side of the equilibrium beam. With the exception of tritium, the magnitude of deflection increases with decreasing \az. 

Protons and $^{3}_{2}$He, the lightest charged fragments, are deflected out of the beam pipe between the first and second dipole magnets. Therefore, to measure these fragments the first detection station must  be placed after the first magnet. Its location is chosen to be at $s=14$\,m. The next station is needed to detect fragments with $2<\az<2.1$. To conform to this deflection pattern, the second station is located at $s=20$\,m from the IP. 
Its goal is to measure the largest \az\ fragments, and for this the detector should be placed as close as possible to the equilibrium beam. Placement of detector elements too close to the circulating beam can cause beam loss, and so to reach higher \az the third station is located further away from the IP where the size of the equilibrium beam becomes smaller and the deflection of the fragments displaces them significantly from the beam. The last station is placed at $s=72$\,m. 

In this calculations the aperture constrains in $x$ direction are taken to be $8\times\sigma$ of the beam in the direction closer to the equilibrium beam.  In the opposite direction they are taken equal to the dimension of the beam pipe, as discussed in Sec.~\ref{sec:tracing}. Detector acceptance in the vertical direction is taken to be $\pm$60\,mm from the  beam center. The aperture of the ZDC is taken to be $\pm$55\,mm~\cite{zdc} at 18\,m from the IP. 

\section{Generators of the spectator fragments}
\label{sec:frag_gen}
Fragmentation of Pb ions in a 158A GeV fixed target experiment was measured by the NA49 Collaboration~\cite{na49:spect} for all fragments, protons, and neutrons.  However, for performance studies of a centrality detector one needs more detailed information about spectator nucleon fragmentation and aggregation.  As discussed in the previous section detecting spectator fragments with \az close to 2.5 is problematic. Such fragments are produced in peripheral collisions (see Figure \ref{fig:Frag_massn} below) and the detector performance should have centrality dependence. To understand this dependence and other factors affecting the detector performance 
spectator fragments were generated by two Monte Carlo generators: \gD and \gQ. 
A comparison between them and to available experimental data is discussed in this section.

The \gD generator is based on Dual Parton Model (DPM) ~\cite{DPM} and is capable of simulating hadron-hadron, nucleus-nucleus, photon-hadron and photon-nucleus interactions from a few GeV up to the highest cosmic ray energies. The \gQ generator is based on the Quark-Gluon String Model~\cite{QGSM} and has the capability of simulating hadron-hadron, nucleus-nucleus and hadron-nucleus interaction. Both models take into account Fermi break up, multi-fragmentation, evaporation, and fission processes.

\begin{figure}[!htb]
\begin{center}
\includegraphics[width=0.5\textwidth]{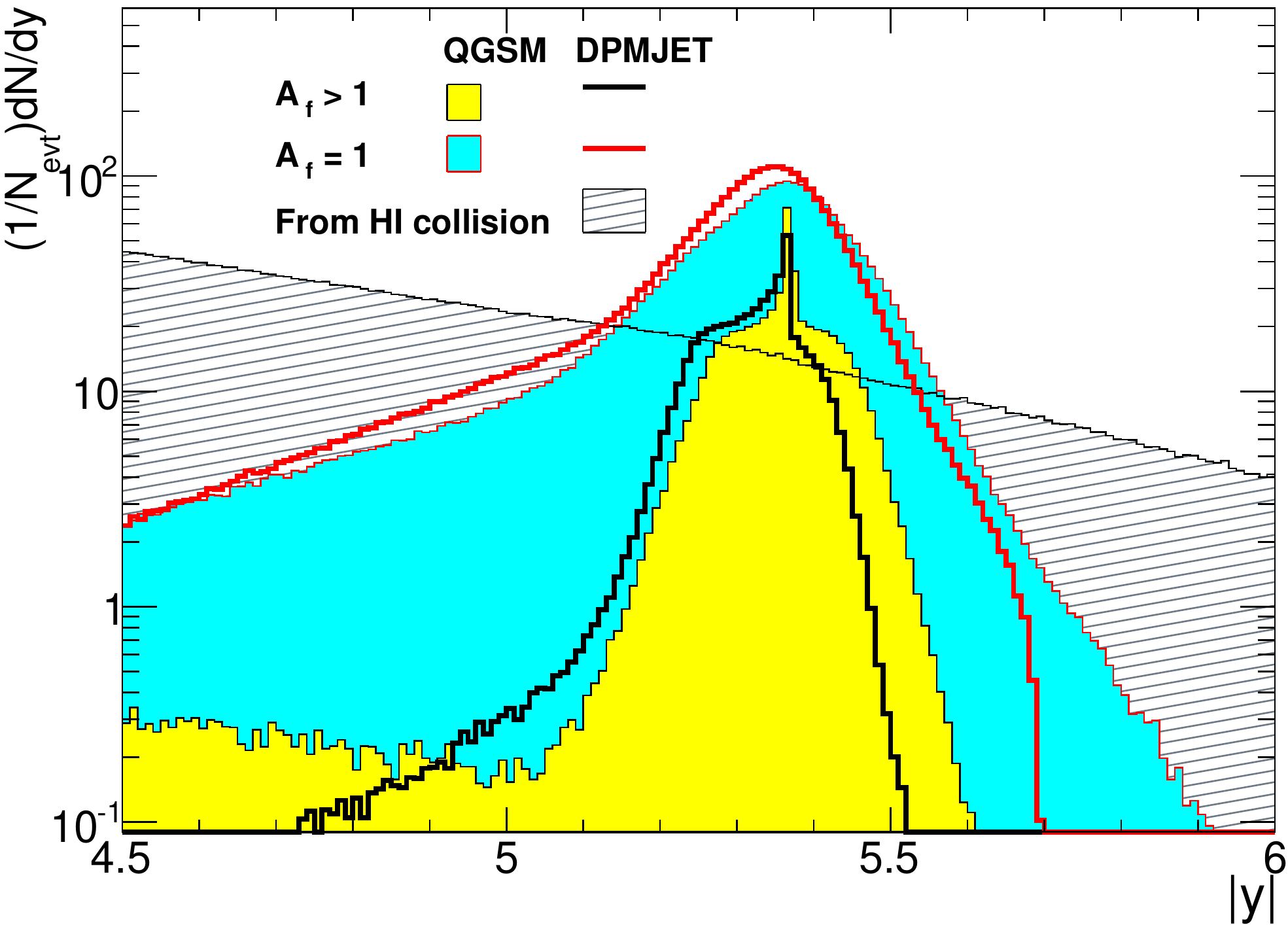}
\caption{The rapidity distribution of the fragments produced per event $1/\Nevt dN/dy$ by the \gQ generator (filled histograms) and by the \gD generator (empty histograms) as a function of rapidity $|y|$. Charged particles produced in HI interaction by the \gD generator are shown with hashed histogram. 
} 
\label{fig:rapidity}
\end{center}
\end{figure}
Both \gD and \gQ generators identify the physics process from which fragments originate and provide the kinematic information for each produced fragment. Spectators have rapidity comparable to the rapidity of the equilibrium beam $y_{\mathrm{beam}}=5.36$. The rapidity distribution of all particles produced by \gD and \gQ generators around the beam rapidity is shown in Fig.~\ref{fig:rapidity}. The results of \gD are shown with empty histograms and the results of \gQ with filled histograms.

Figure~\ref{fig:rapidity} shows the rapidity distribution of final state particles in the forward region near the beam rapidity.  
Final state particles with non-zero baryonic number and $|y|>5$, generated in a physics process that involves only one of the colliding ions are selected as spectators.  Spectator fragments with $\af>1$ have a distinct peak at the rapidity of the beam due to heavier particles sharing a very similar trajectory with the equilibrium beam. Spectators with $\af=1$ (protons and neutrons) have a wider distribution. 
Charged particles produced in HI interactions, \textit{i.e.} non-spectator particles, including baryons, are also shown in the plot.

\begin{figure}[!htb]
\begin{center}
\includegraphics[width=0.5\textwidth]{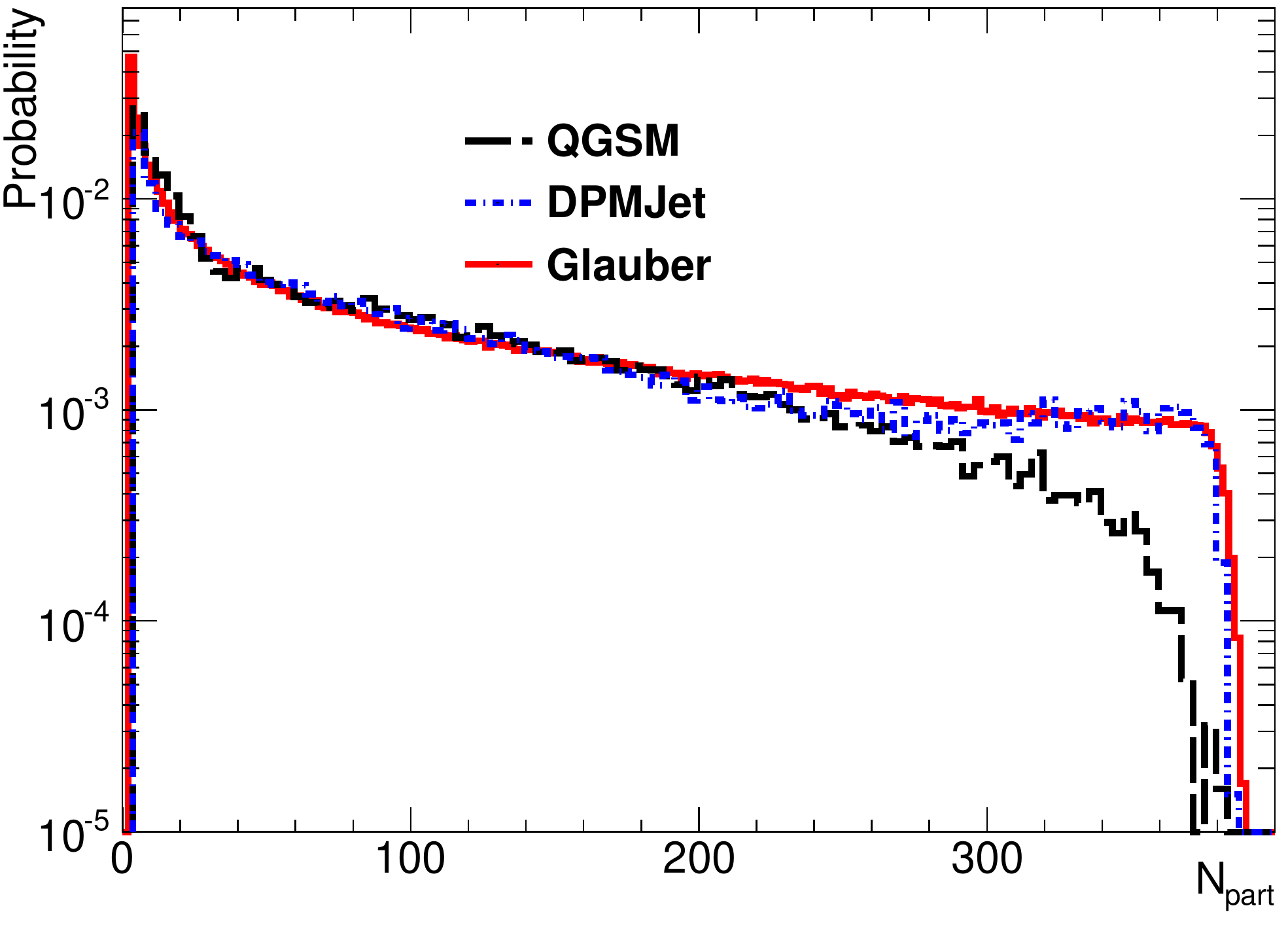}
\caption{The \Npart distribution produced by the \gD and the \gQ generators and by the MC Glauber model.} 
\label{fig:Npart}
\end{center}
\end{figure}

After accounting for all the spectators produced in each modelled event, \Npart is defined according to Eq.~\ref{eq:npart}.  Figure~\ref {fig:Npart} shows the \Npart probability distribution generated by the \gD and the \gQ models as well the Glauber model. 
The \gD and Glauber models show good agreement whereas the \gQ model shows a significant deficit at high \Npart\ and excess at lower \Npart.

Spectator fragmentation into final state particles within a given \Npart\ window is also found to be different between the \gD and \gQ generators. The difference is seen at all centralities in Fig.~\ref{fig:Frag_massn} which shows the \af distribution of spectator fragments produced by the generators for different \Npart intervals.
\begin{figure}[!htb]
\begin{center}
\includegraphics[width=0.5\textwidth]{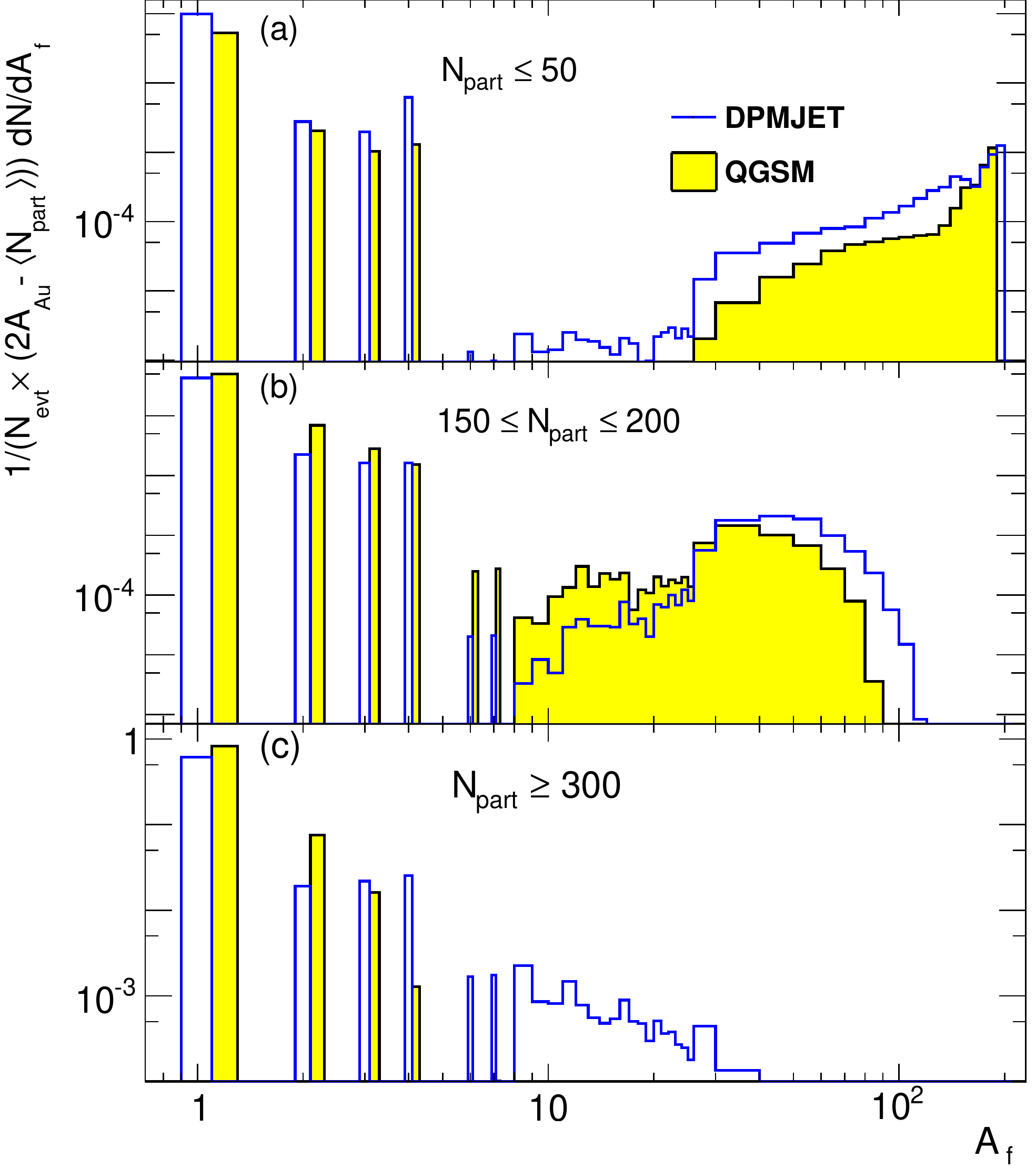}
\caption{The \af distribution produced by the \gD (line) and by the \gQ (filled histogram) generators for the events with different \Npart. The distributions are normalized by the number of MC events and by the average number of spectator nucleons. Values in the bins $\af<8$ are staggered from the centre of the bin.} 
\label{fig:Frag_massn}
\end{center}
\end{figure}
The distributions show that \gQ\ tends to produce lighter fragments than \gD in central and mid-central events, and in peripheral events the trend is opposite. 

\begin{figure}[!htb]
\begin{center}
\includegraphics[width=0.5\textwidth]{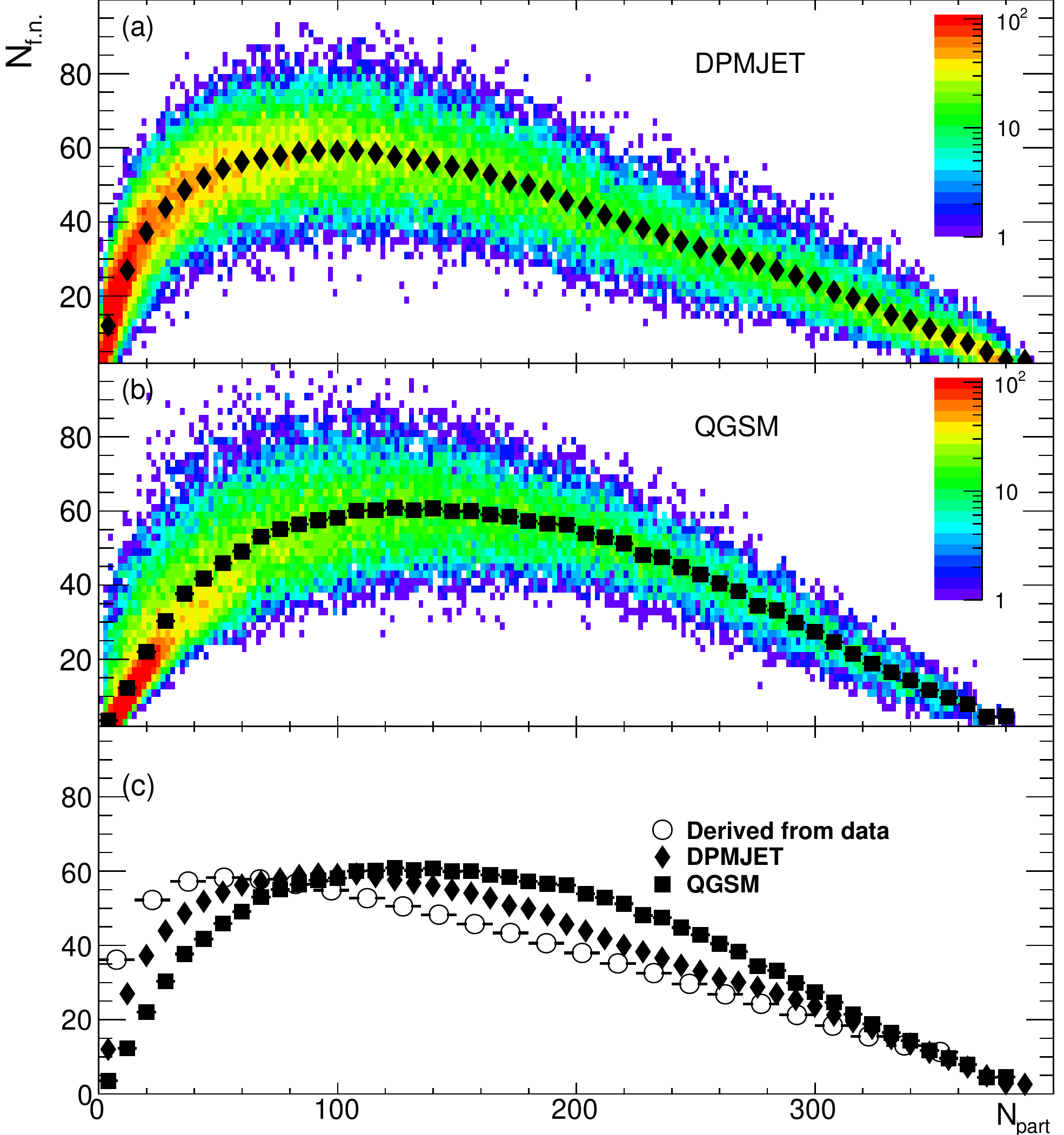}
\caption{The number of free spectator neutrons within ZDC acceptance versus \Npart. Panel (a) shows the results of the \gD generator and panel (b) of the \gQ generator. The markers in the plot are the averaged values.  In panel (c) the two curves from panel (a) and (b) are compared with the same quantity derived from the data~\cite{BBC_ZDC} as described in the text.} 
\label{fig:ZDC_BBC}
\end{center}
\end{figure}

Although the full fragmentation spectrum has not been measured at RHIC, the performance of the two models may be gauged by comparing their production of free spectator neutrons, \nf, (\textit{i.e.} a fragment composed of a single neutron) to data measurements. 
This can be done using the response of the ZDC calorimeters installed in the RHIC experiments, which measure the energy carried mainly by free spectator neutrons. 
Fig.~\ref{fig:ZDC_BBC} shows the
\nf as a function of \Npart for the generators used in the analysis. Panel (a) shows the result for \gD and panel (b) for \gQ. The filled symbols superimposed on the scatterplot are the mean values of \nf  at each value of \Npart. 

These values are compared to the values derived from the data published by the PHENIX experiment~\cite{BBC_ZDC}. The distribution of energy measured in the acceptance of PHENIX ZDC versus charge measured by the Beam-Beam Counters (BBC) is shown in Fig.1 of Ref.~\cite{BBC_ZDC}. The values on the axes of the plot are given in arbitrary units. The charge measured in the BBC is proportional to the number of produced particles which is proportional to \Npart. This relation is used by the PHENIX experiment to determine centrality. One can further approximate that the maximum value of the BBC, equal to $\sim1.5$ in arbitrary units of Fig.1 in Ref.~\cite{BBC_ZDC} corresponds to the maximum number of participants $\Npart=353$ in Table~XIII of the same reference. The dominant part of the energy measured by the ZDC is carried by the free neutrons, each delivering on average the same energy - 100\,GeV. Therefore, the ZDC response is proportional to \nf. The maximum averaged \nf in the data is assumed to be the same as generated by \gD and \gQ within the aperture of the ZDC. 

Panel (c) of Fig.~\ref{fig:ZDC_BBC} shows the comparison between the \nf extracted from the data and that produced by the generators. The \gD generator better describes the centrality dependence of \nf compared to \gQ. In the interval \Npart above 150, the curve produced by the \gD generator shows the same trend as the data estimate. The absolute values are different, which can be an artefact of the procedure used to derive the data estimate. At low \Npart both generators show significant deviations from the estimate, producing lesser \nf. The implication of this discrepancy is discussed in Sec.~\ref{sec:concl}.


\section{Detector performance}
\label{sec:det_perf}
The detector performance depends on how completely and accurately spectator fragments can be reconstructed based on their kinematic reconstruction in the detectors.  Several factors listed below have an impact on the measurement of spectators.

\subsection{Collider effects} 
Ions in the beam have spatial and angular dispersions defined by the $\beta$ function of the collider and the emittance of the beam. 
Most of the modern collider based detectors are equipped with a vertex detector which can determine the position of an event vertex in transverse plane with an accuracy better than the coordinate dispersion of the particles in the beam. However in these studies, precision vertex information is not used and the transverse dispersions present in the equilibrium beam smear the calculation of spectator kinematics
, see Eq.~\ref{eq:matrix}. The magnitude of this effect is visible in panel (b) of Fig.~\ref{fig:dev_physproc}.

Background hits in the detector stations produced by the particles outgoing from the equilibrium beam, or by their secondaries, are not considered in this work. These may be coming from the beam-gas interactions, from the interactions of spectator fragments hitting the walls of the beam pipe or collider structure elements. They can produce significant number of hits in the stations and affect the detector performance. However, understanding of these processes requires more realistic simulation of the collider structure elements and of the detector hardware inside the stations, which lies outside the scope of this paper. The pile-up, caused by multiple HI interactions inside the same crossing of the ion beams or coming from two subsequent beam crossings is also not considered. 

\subsection{Collision effects} 

Particles created in HI collisions with sufficiently high rapidity form a background to the spectators in the detector.  These are simulated by \gD\ and are traced though the collider structure in the same way as spectator fragments. In the \gQ generator produced particles are not simulated. 

The detector performance is directly related to the ability with which \af/\zf\  and ultimately \af\ can be reconstructed.   
The dominant factor that weakens the correlation between the particle's position in the detector and its \af/\zf\ value is the Fermi motion of the nucleons inside the colliding ions. In the process of fragment creation it results in an angle of the spectator fragments with respect to the initial direction of the ion and changes its longitudinal momentum. In these studies the effects of the Fermi motion in the longitudinal and transverse directions are taken as modelled by the generators, but they are shown separately.

Let \pf be the Fermi momentum of a nucleon in the ion rest frame, then 
in the laboratory frame the average angle of a fragment with respect to the ion direction 
and the longitudinal momentum dispersion relative to \pz are given by Eq.~\ref{eq:fermi}:
\begin{eqnarray}
\langle y\prime\rangle = \langle x\prime\rangle\approx \frac{1}{\sqrt{3\af}}\frac{p^{F}}{\pz} \nonumber \\
\langle\Delta p_{z}/p_{z}\rangle \approx \frac{1}{\sqrt{3\af}}\frac{\pf}{m_{\mathrm{N}}}
\label{eq:fermi}
\end{eqnarray}
where $m_{\mathrm{N}}$ is nucleon mass. Both dispersions 
decrease $\propto1/\sqrt{\af}$, however the angular dispersion of fragments also diminishes with beam energy ($\pz \approx \sqn/2$), whereas the longitudinal momentum dispersion does not depend on the beam energy.


\subsection{Spectator deflection in the detector stations} 
\label{sec:stations}

\begin{figure}[!htb]
\begin{center}
\includegraphics[width=0.5\textwidth]{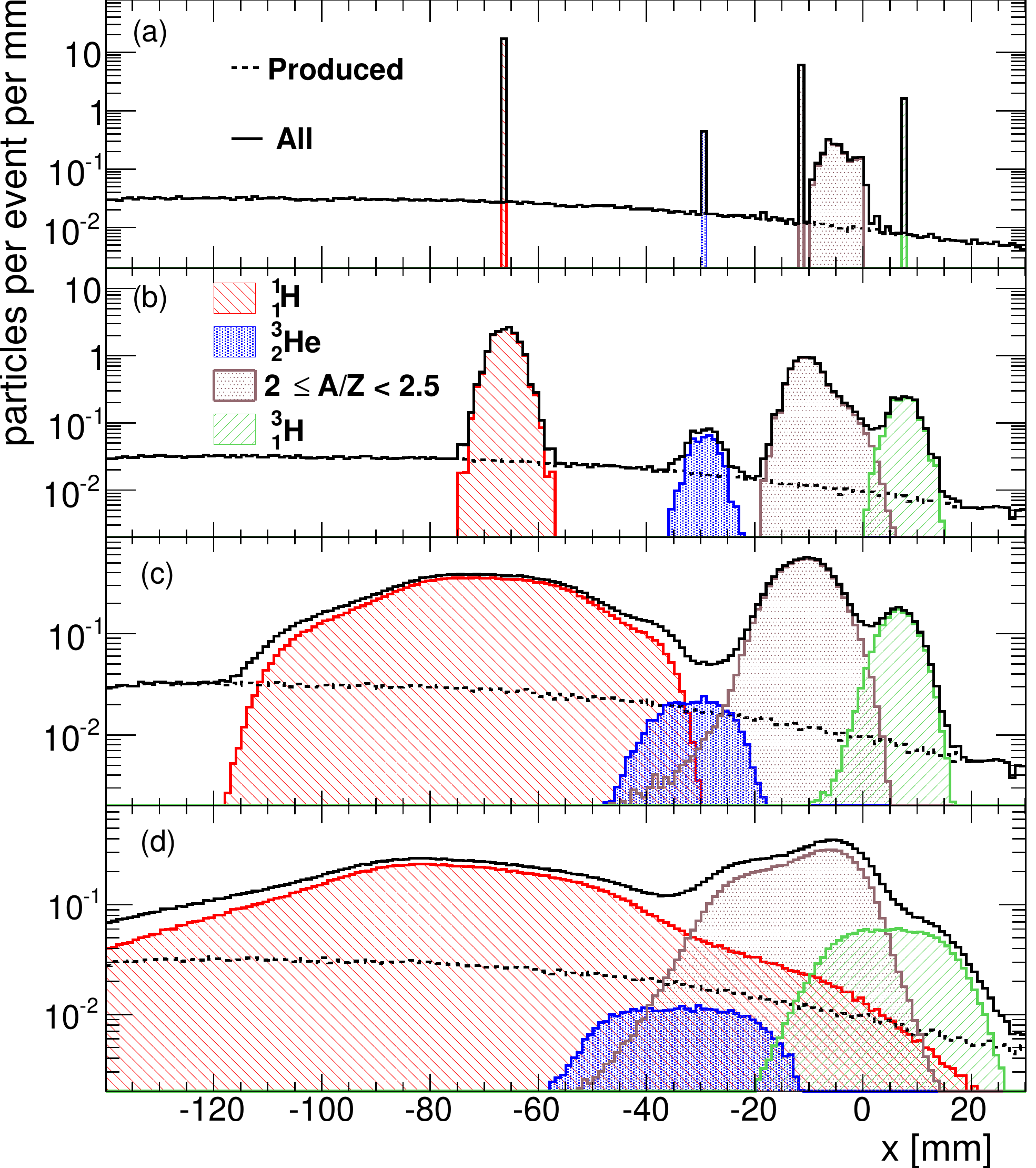}
\caption{Position of different types of charged particles at the location of the first detector station ($s=14$\,m). Relative amplitudes of different particles correspond to the sample of all centralities, generated with \gD. Panel (a) shows the "ideal" case. Panel (b) includes into consideration the dispersions in the ion beam. Panel (c) includes the ion beam dispersions and the longitudinal Fermi motion and panel d) includes the beam dispersions and the full Fermi motion. 
} 
\label{fig:dev_physproc}
\end{center}
\end{figure}

The positions of charged spectator fragments in the $x$ direction at the location of the first detector station ($s=14$\,m), calculated using Eq.~\ref{eq:matrix}, are shown in Fig.~\ref{fig:dev_physproc}. 
In the case that a single fragment enters the aperture of more than one station it is considered to be measured in the station closest to the IP and is ignored in the subsequent stations. 

Panel (a) corresponds to the ``ideal" case, in which the charged fragment distributions are calculated without any distortions. 
Peaks from left to right correspond to protons, $^{3}_{2}$He, particles with $\az=2$, $2<\az<2.5$, and tritium (which appears in the positive region of the axis).  Integrals of the peaks correspond to the production rates of spectator fragments produced in all centralities. In this ideal case, all spectator fragments with the same \az\ arrive at the same point in the detector resulting in sharp peaks. The equilibrium beam, with $A/Z=2.5$, arrives at  $x=0$ and is not shown in the figure.  
Results of calculations taking into account  angular and spatial dispersions of the ion beam are shown in panel (b). This is done by assigning to each spectator fragment position and angle of the corresponding ion at the IP. 
Including the longitudinal Fermi motion component makes the peaks significantly wider and they start to overlap as shown in panel (c). Adding to these effects also the transverse components of the Fermi motion is shown in panel (d) of the figure. 

The beam dispersions and the dispersions due to the transverse Fermi motion depend on the parameters of the collider. At LHC energies transverse Fermi motion plays a less significant role than at RHIC, while at the NICA collider~\cite{NICA} their contributions are more significant. The longitudinal component remains the same at all energies. 

\begin{figure}[!htb]
\begin{center}
\includegraphics[width=0.5\textwidth]{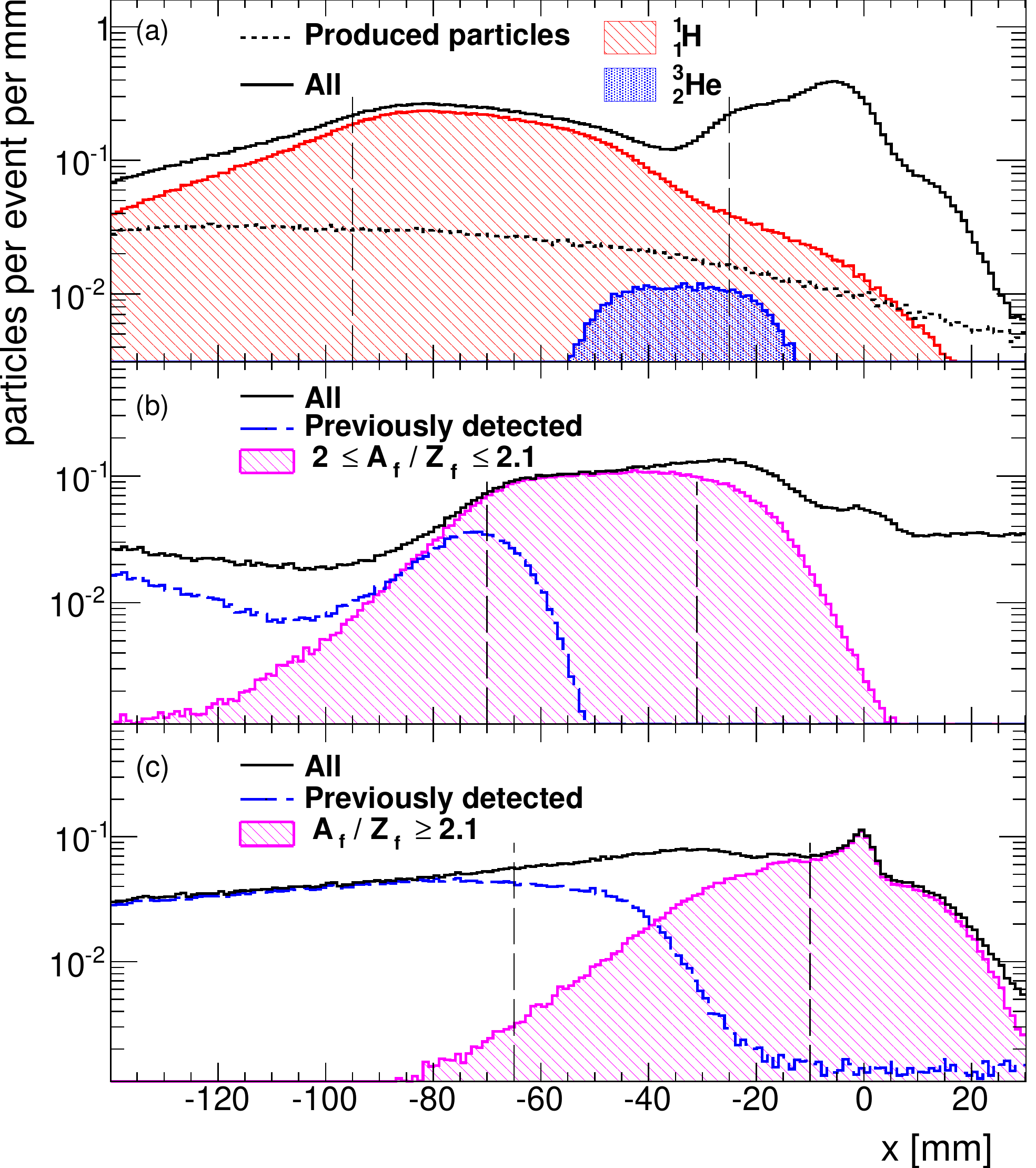}
\caption{Position of different types of charged particles at the locations of the first station (a),  the second station (b) and the third station (c). Stations are optimised to measure particles shown with hashed histograms. Distributions of spectators detected the stations closer to the IP are shown with dashed histograms. The station acceptance is shown with vertical lines. The distributions in the third stations are inverted: $x\rightarrow-x$.} 
\label{fig:def_d1d2d3}
\end{center}
\end{figure}

Figure~\ref{fig:def_d1d2d3} shows the $x$ position distributions of charged spectator fragments in each of the three detector stations.
The calculations are performed using the \gD generator, including all relevant effects discussed above. Panel (a) shows the deflection in the first station at $s=14$\,m,  panel (b) the second station at $s=20$\,m, and panel (c) the third station at $s=72$\,m. The $x$ positions of the particles in the third stations are inverted $x\rightarrow-x$ because it is located on the other side of the equilibrium beam than the other two stations as shown in Fig.~\ref{fig:tracing}.

The primary goal of the first station is to detect protons and $^{3}_{2}$He spectators. The distributions shown in panel (a) are the same distributions as the proton, $^{3}_{2}$He, and inclusive distributions shown in panel (d) of Fig.~\ref{fig:dev_physproc}. Spectators entering into the detector aperture, shown with vertical bars, are considered detected.  
The second station is intended to detect fragments with $2\leq\az\lesssim2.1$. The distribution of such particles at the location of the second station is shown with hashed histogram in panel (b). Spectators residing within the station aperture are detected in the second station. Spectators detected in station 1 are shown by the dashed histogram. The third stations is designed to measure spectators with $\az\gtrsim2.1$. Their distribution at the location of the third station is shown in panel (c) with hashed histogram. Spectators detected in the other two stations are shown by the dashed histogram. 

\begin{figure}[!htb]
\begin{center}
\includegraphics[width=0.5\textwidth]{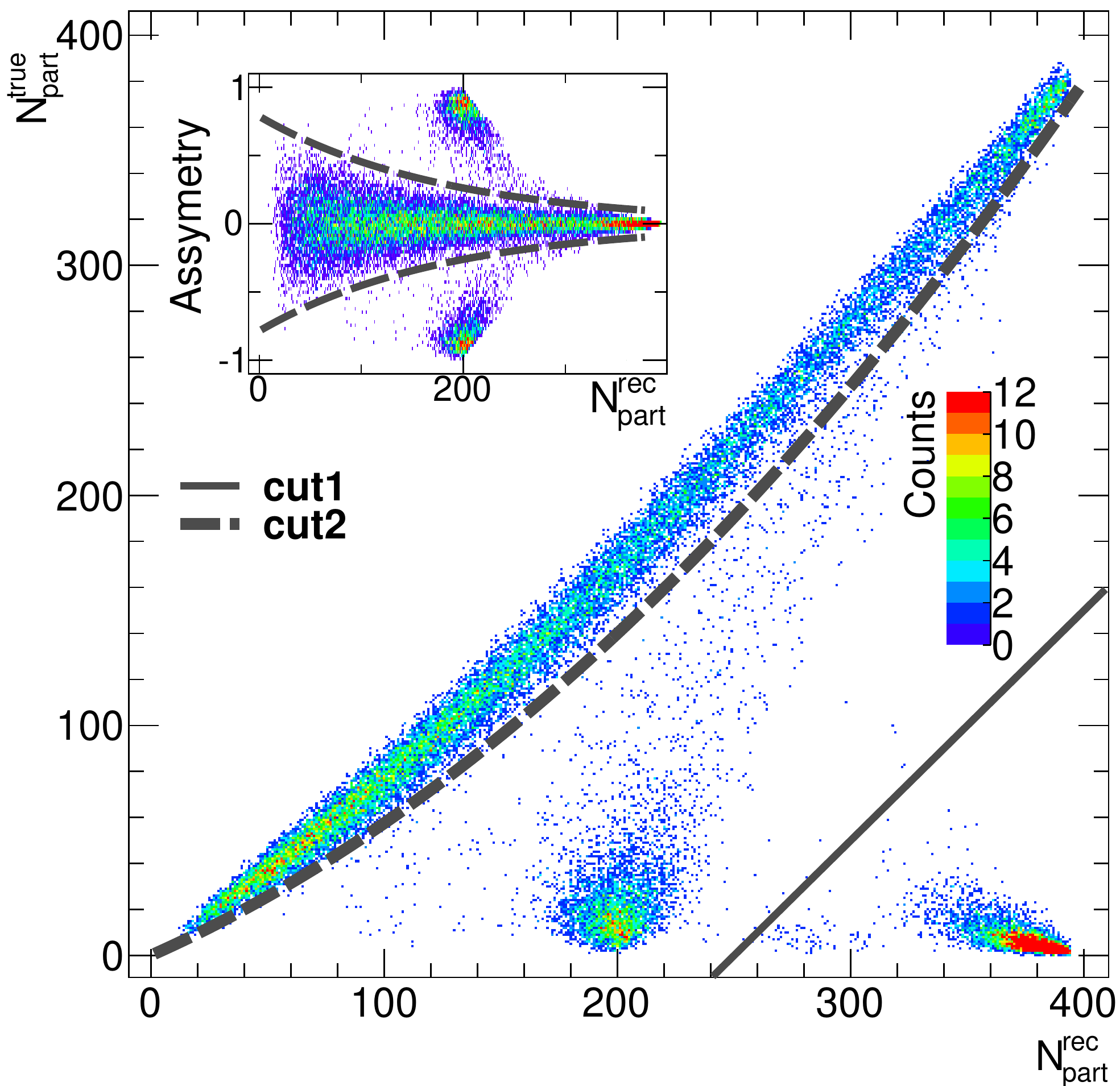}
\caption{The true number of participants, \Npt versus the number of reconstructed participants, \Npr. The insert shows the asymmetry in the number of participants reconstructed on both sides of the IP. Lines indicate event selection criteria explained in the text.} 
\label{fig:recoNpart_trNpart}
\end{center}
\end{figure}

To quantify the performance of an ideal detector, the assumption is made that all detected spectator fragments are reconstructed with their true \af.  Then using Eq.~\ref{eq:npart} a reconstructed number of participants, \Npr, is calculated, which can be compared to the event's true number of participants, \Npt.  The calculation of \Npr\ includes the contribution of background particles produced in the HI collision that enter the detector.  Figure \ref{fig:recoNpart_trNpart} shows the two-dimensional \Npt versus \Npr distribution calculated using \gD.

There are three distinct regions in the plot. The region at ($\Npr\gtrsim 320$, $\Npt\lesssim40$) corresponds to events in which  
 two heavy fragments with \az close to 2.5 are produced in a peripheral collision and neither is reconstructed by the detector. In such events the centrality cannot be determined, however such events should have a significant mismatch between the large \Npr measured by the centrality detector and low number of produced particle measured by any other detector subsystem. Such an identification procedure is equivalent to removing events below the solid line shown in the figure.

The region at ($\Npr\sim 200$, $\Npt\lesssim60$) corresponds to peripheral events in which one heavy fragment escapes detection.
These events can be identified by comparing the response of the centrality detectors on both sides of the IP. The asymmetry in the number of participants, $(N_{\mathrm{part}}^{N}-N_{\mathrm{part}}^{S})/(N_{\mathrm{part}}^{N}+N_{\mathrm{part}}^{S})$ where $N$ and $S$ are the opposite sides of the IP, is shown in the insert. Rejecting events with high asymmetry as indicated by the dashed line in the insert would result in rejecting events located below the dashed line in the main area of the figure. The centrality in this class of events can still be measured on one side and extrapolated to a total \Npart by multiplying by a factor of 2.  In the presented analysis the extrapolation is not done and these events are not further considered. 

The bulk of events are close to the diagonal \Npt=\Npr. These are the events with properly reconstructed \Npart.  Figure~\ref{fig:recoNpart_trNpart} shows that even in this region, \Npr\ somewhat underestimates \Npt\ due to fragments which miss the detectors.  This necessitates a correction for a finite acceptance of the detector stations, however this correction should be based on data, and does not require a model.


\section{Results}
\subsection{Efficiency and resolution of centrality determination} 
\label{sec:res_main}
\begin{figure}[!htb]
\begin{center}
\includegraphics[width=0.5\textwidth]{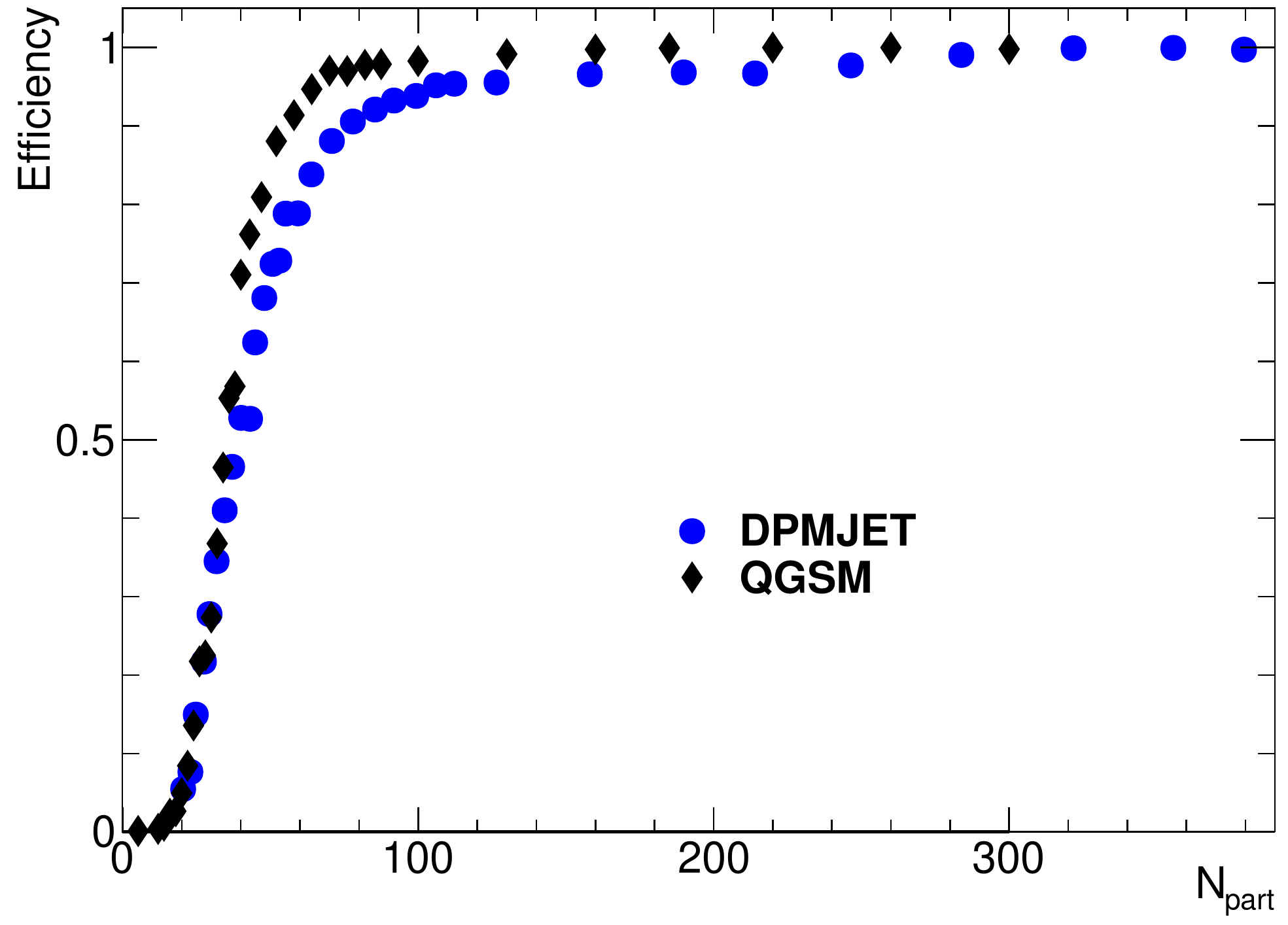}
\caption{Fraction of all events in which the \Npart determination is possible, the centrality determination efficiency.} 
\label{fig:eff}
\end{center}
\end{figure}

The fraction of all events in which centrality can be determined, the centrality determination efficiency, is shown in Fig.~\ref{fig:eff} as a function of \Npart. 
The two curves correspond to the results of the \gD and \gQ generators and both  approach unity at high \Npart. In peripheral events, the efficiency rapidly falls to zero with decreasing \Npart. 
The low efficiency notwithstanding, low \Npart\ events which are detected have a robust centrality determination even in $\Npr<20$.
In peripheral events the \gD generator based calculations show higher efficiency for the same \Npr compared to \gQ. This is related to the differences in the \af distributions discussed in Sec.~\ref{sec:frag_gen}. 

\begin{figure}[!htb]
\begin{center}
\includegraphics[width=0.5\textwidth]{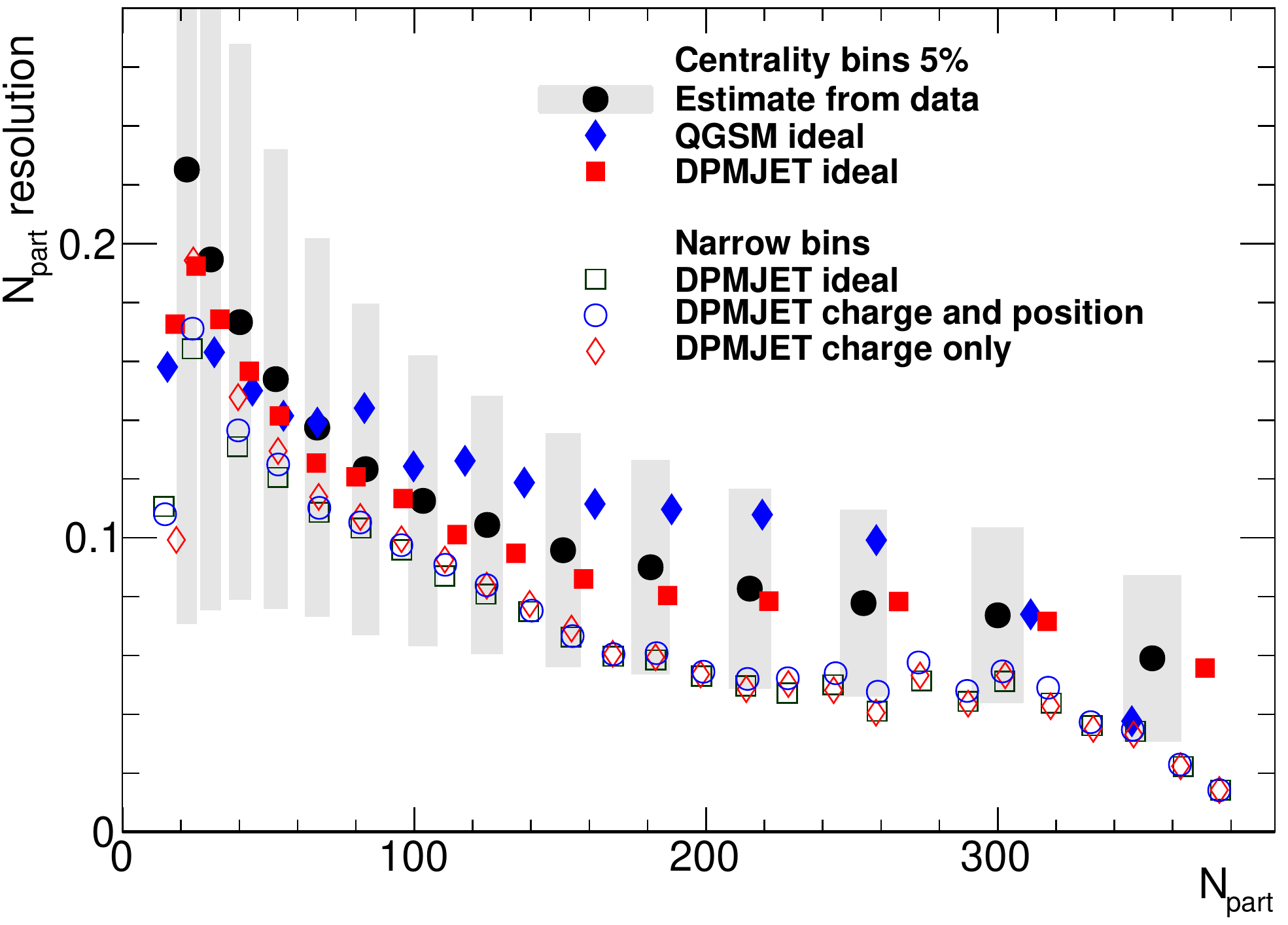}
\caption{Resolution of \Npart determination for calculations using the \gD and the \gQ generators in different centrality bins (filled markers) and for different choices of detector technologies (open markers). Calculated resolution is compared to the estimation based on the data, which is derived from publications~\cite{BBC_ZDC,Glauber_res}.} 
\label{fig:res}
\end{center}
\end{figure}

The resolution of the \Npart determination is defined as the R.M.S. of the \Npt distribution for a given \Npr  divided by its mean value $\langle\Npt\rangle$. Resolution of \Npart is shown in Fig.~\ref{fig:res}. 
The width of the \Npart distribution has two contributions. The first depends on the width of the centrality interval, i.e. the width of the percentile (or \Npr) over which the averaging is done. The second is the intrinsic resolution of the method and the detectors that are used for measuring \Npart. Figure~\ref{fig:res} has two sets of curves: filled markers correspond to the resolution in predefined centrality intervals, open markers correspond to the intrinsic detector resolution. 

The results of calculations are compared to an estimate based on the data published by the PHENIX experiment which is shown with filled circles. The estimate is based on the widths of the \Npart distributions in 5\% centrality intervals shown in the left panel of Fig.18 in Ref.~\cite{Glauber_res}. The values in the plot are divided by \avgNpart in the same centrality intervals given in Ref.~\cite{BBC_ZDC} and are plotted versus \avgNpart. The error bars correspond to the systematic uncertainties which are given in the same publication. The results of the calculations for \gD and \gQ models are shown in the same centrality intervals as the data estimate, calculated using \Npr. These estimates include the width of the centrality intervals and the intrinsic resolution of the method, but not the resolution of the detector, which is discussed below. As one can see the resolution depends on the choice of  generator and is comparable to currently used techniques. 

To address the question of intrinsic resolution of the method and the contribution which is coming from possible choice of detector technology to be used in the detector stations, the resolution was calculated with the \gD generator in narrow \Npr intervals. The resulting curves are shown in Fig.~\ref{fig:res} with open markers. Open squares corresponds to the case when each \af is measured perfectly, i.e. the true \af is accepted for each detected particle. The open circles (``charge and position") correspond to the case in which the \zf of the fragment is measured perfectly, but the mass is taken at an average value of all spectators with measured \zf at the $x$-position in the detector. The open diamond (``charge only") markers correspond to the case when the coordinate is not reconstructed at all, but the mass is taken as an average mass of all fragments for a given \zf. 

The curves are all similar, because the dominant factor which determines the resolution is the loss of spectator fragments coming from increased deflection due to Fermi motion (see Fig. \ref{fig:dev_physproc}). 
The curves are not flat at $\Npart\approx 250$, this is an artifact of the asymmetry cut shown in the insert of Fig.~\ref{fig:recoNpart_trNpart}. 

\subsection{Possible choices of the detector technology} 
\label{sec:res_det}
The key requirement for each detector station is the ability to reconstruct \zf.   
A suitable choice to achieve this using existing technology is a Cherenkov radiation detector. The resolution needed to distinguish two fragments with charges \zf-1 and \zf\ is estimated by Eq.~\ref{eq:q_res} 
\begin{equation}
\frac{dq}{q} \lesssim \frac{\langle Z_{f}^{2}-(Z_{f}-1)^{2}\rangle}{\sqrt{12}\langle Z_{f}^{2}\rangle}\approx \frac{\langle Z_{f}\rangle}{\sqrt{3}\langle Z_{f}^{2}\rangle}\approx \frac{1}{\sqrt{3}\langle Z_{f}\rangle},
\label{eq:q_res}
\end{equation}
neglecting the difference between $\langle Z_{f}^{2}\rangle$ and $\langle Z_{f}\rangle^{2}$.  

Different stations are designed to register particles with different \az and therefore different \zf, as explained in Sec.~\ref{sec:stations}. 
Station 1 mainly detects fragments with $\zf=1$ and 2. The mean charge of fragments in station 2 is $\langle Z_{f}\rangle=20$ and is $\langle Z_{f}\rangle=40$ in station 3. From Eq.~\ref{eq:q_res}, the required resolution for measuring \zf in each station  is 30\%, 3\% and 1.5\% respectively. 
A Cherenkov detector with a 5\,cm radiator, an index of refraction in the range of optical glass, 20\% light collection efficiency, and 10\% photosensor quantum efficiency yields approximately 20 photoelectrons per fragment in station 1, $\sim4\times10^{3}$ in station 2, and $\sim1.5\times10^{4}$ in station 3.  This would provide enough photoelectrons to meet the desired resolution. The Cherenkov detector must have sufficient granularity to measure multiple fragments simultaneously. The average numbers of fragments in each stations does not exceed 10, suggesting that the detector has to have from tens to a hundred individual channels.

Figure~\ref{fig:res} shows that measuring fragment positions has only small effect on the final \Npart resolution.  However, measurement of the fragment position is important for detector alignment and for rejecting background, it can be useful to trace particles from one station to another. A possible choice of detector technology for determining fragment position is a silicon pixel based tracker with several layers along the fragment trajectory. A similar choice of detector technologies is suggested for the forward physics upgrade of the ATLAS detector at the LHC~\cite{atlas-afp}.

\subsection{Measurement of the event plane orientation} 
\label{sec:ep}

Azimuthal anisotropy of particle emission in heavy-ion collision is an important observable to understand the medium created in HI collision. The harmonics of the azimuthal anisotropy of particle emission are studied by all HI experiments~\cite{PhysRevLett.96.032302,PhysRevC.85.064914, atlas_flow,alice_flow,PhysRevC.89.044906}. The measurement of the $n$-th harmonic relies on the determination of particle emission angles with respect to the event plane $\Psi_{n}$ of the corresponding harmonic. The event planes are measured in the forward region using particles produced in the collision, except for $\Psi_{1}$, which cannot be determined with produced particles and is measured using the ZDC.

The proposed detector offers an opportunity to measure $\Psi_{1}$. Determination of the $\Psi_{1}$ event plane can be made by measuring spectator fragment positions in all three stations. The resolution $d\Psi_{1}$ can be then estimated as: 
\begin{eqnarray}
\label{eqn:psi_res}
d\Psi_{1} = \frac{\langle\sigma\rangle}{\theta} \langle|f(\Psi)|\rangle
\label{eq:dpsi}
\end{eqnarray}
where $\langle\sigma\rangle$ is the average emission angle of all spectators, and $\theta$ is the relative angle between the direction of spectator fragments and the ion ($\theta$ is assumed to be the same for all fragments in an event).
The factor $f(\Psi)$, accounts for resolution differences in the $x$ and $y$ directions.
In the case $\sigma_x \approx \sigma_y$, the modulus of this function averaged over all angles is $\approx 1$. This condition is true in all three detector stations. The average emission angle $\langle\sigma\rangle$ can be estimated by measuring deflection of the particles in the detector stations.
\begin{eqnarray}
\langle\sigma\rangle = \frac{\sum_{d}\sum_{i}\frac{x_{i}-\langle x_{i}\rangle}{a_{1,2}(d)}A_{f}^{i}}{\sum_{d}\sum_{i}A_{f}^{i}}
\end{eqnarray}
The index $i$ refers to spectators in a given detector station and $d$ to the three detector stations.  It can also be done in each station individually. The ($x_{i} - \langle x_{i}\rangle$) is relative deflection of the fragment in a station with respect to an average position of all fragments with the same \az. Coefficient $a_{1,2}(d)$ is the matrix element for $d$-th station in Eq.~\ref{eq:matrix}. Each spectator fragment is summed with a weight equal to the \af, to account for the fact that heavier fragments have lesser distortion due to Fermi motion, and therefore their contributions to the $\Psi_{1}$ measurement are more accurate.

\begin{figure}[!htb]
\begin{center}
\includegraphics[width=0.5\textwidth]{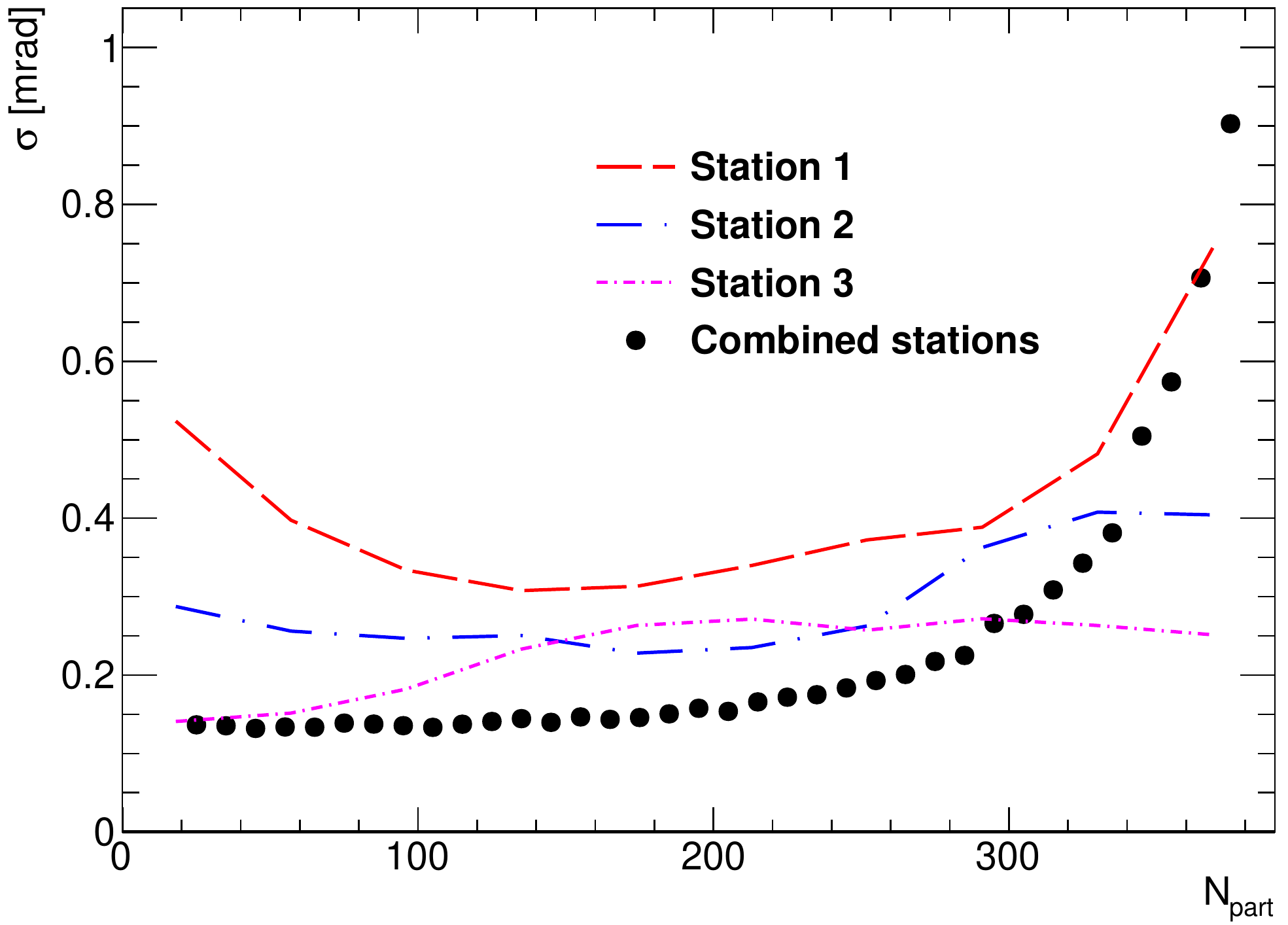}
\caption{Spectator emission angle determination accuracy $\langle\sigma\rangle$ as a function of \Npart for each station and for the combination of all three stations.} 
\label{fig:rxnp_res}
\end{center}
\end{figure}

Figure~\ref{fig:rxnp_res} shows $\langle\sigma\rangle$ as a function of \Npart. Lines correspond to the results of measuring $\langle\sigma\rangle$ with individual stations on both sides of the IP. 
Markers correspond to the combination of all three stations. The combined result does not include an additional measurement which can be provided by the ZDC. As measured by the ALICE collaboration the average deflection of neutrons in  the ALICE ZDC for 30--40\% centrality is 0.92\,mm at 110\,m \cite{PhysRevLett.111.232302}. Assuming that the angle is inversely proportional to \sqn, Eq.~\ref{eq:dpsi} and Fig.~\ref{fig:rxnp_res}, it follows that $d\Psi_{1}\approx1.1$\,rad. 

\section{Conclusions}
\label{sec:concl}
This paper presents a detector concept for the direct measurement of the number of participants in heavy ion collisions by detecting spectator fragments. The performance of the detector is evaluated based on the example of \AuAu interactions at \energy in RHIC. The location of 3 detector stations, integrated into the RHIC structure, are optimised for the best detector performance. The main performance parameters, such as the efficiency of centrality determination and resolution in measuring the number of participants is presented as a function of collision centrality based on the fragmentation modelled by the \gD and \gQ generators. 

The detector performance is compared to present techniques for measuring centrality and is found to be comparable to them. The results are significantly different for the \gD and \gQ generators which have different distributions of produced fragments for the same \Npart. Comparison of generators to the existing data is limited, and shows that both generators have significant deviations from the measured quantitates and that the \gD better reproduces available data.

Modern detector technologies are shown to be adequate to perform the measurements. The proposed concept offers an opportunity to make a precise measurement of the orientation of the first order event reaction plane. The main advantage of the centrality detector is in measuring the number participants in a model independent way, with no correlation to produced particles.

\section{Aknowledgements}
\label{sec:akn}
The authors are thankful to our colleagues, Prof. Itzhak Tserruya and Dr. Ilia Ravinovich at the Weizmann Institute of Science for numerous useful discussion and help in preparing the document. Authors express their gratitude to Prof. Nestor Armesto from University of Santiago de Compostela, Dr. Oleg Rogachevsky from the Joint Institute of Nuclear research at Dubna for ideas and help in using particle generators. Authors are thankful to Dr. Vadim Ptitsyn from the Brookhaven National Laboratory for his help with the MAD-X code and information about RHIC structure. Work of Dr. S.~Tarafdar is supported by "VATAT Program for Fellowships for Outstanding Post-doctoral Researchers from China and India".






\bibliographystyle{elsarticle-num}

\end{document}